\documentclass[12pt]{iopart}
\usepackage{iopams}  
\usepackage{epsfig}
\usepackage{graphicx}

\begin{document}

\title[Trento Workshop Summary]{Initial state fluctuations and final state correlations: Status and open questions}
\author{Andrew Adare${}^{1}$, Matthew Luzum${}^{2,3,4}$ and Hannah Petersen${}^{5,6}$\\[.4cm]}

\address{
${}^1$~Yale University, Department of Physics, New Haven, CT 06520, United States\\
${}^2$~Institut de physique th\'eorique de Saclay (CNRS URA2306), F-91191 Gif-sur-Yvette, France\\
${}^3$McGill University,  3600 University Street, Montreal QC H3A 2TS, Canada\\
${}^4$Lawrence Berkeley National Laboratory, Berkeley, CA 94720, USA\\
${}^5$~Department of Physics, Duke University, Durham, North Carolina
27708-0305, United States\\
${}^6$~Frankfurt Institute for Advanced Studies, Ruth-Moufang-Strasse 1, 60438 Frankfurt am Main, Germany}

\ead{aadare@cern.ch}
\ead{mluzum@physics.mcgill.ca}
\ead{petersen@fias.uni-frankfurt.de}

\begin{abstract}
The recent appreciation of the importance of event-by-event fluctuations in relativistic heavy-ion collisions has lead to a large amount of diverse theoretical and experimental activity.  In particular, there is significant interest in understanding the fluctuations in the initial stage of a collision, how exactly these fluctuations are propagated through the system evolution, and how they are manifested in correlations between measured particles.
In order to address these questions a workshop was organized on ``Initial State Fluctuations and Final State Correlations'', held at ECT* in Trento, Italy during the week of 2--6 July, 2012.   
The goal was to collect recent work in order to provide a coherent picture of the current status of our understanding, to identify important questions that remain open, and to set a course for future research.  Here we report the outcome of the presentations and discussions, focusing on the most important conclusions.
\end{abstract}

%Uncomment for PACS numbers title message

\pacs{25.75.-q,25.75.Ag,24.10.Lx,24.10.Nz}

% PACS Relativistic heavy-ion collisions, Global features in relativistic heavy
%ion collisions, Monte Carlo simulations (including hadron and parton cascades
%nd string breaking models),  Hydrodynamic models 
% Keywords required only for MST, PB, PMB, PM, JOA, JOB? 
%\vspace{2pc}
%\noindent{\it Keywords}: Article preparation, IOP journals
% Uncomment for Submitted to journal title message
%\submitto{\JPA}
% Comment out if separate title page not required

\maketitle

\section[Introduction]{Introduction}
\label{intro}
Relativistic heavy-ion collisions probe the extreme high-temperature regime of the strong
interactions. Specifically, the main goal is to create and characterize a deconfined state of matter, the quark-gluon plasma (QGP)â and investigate the nature of a deconfinement phase
transition. The Relativistic Heavy Ion Collider (RHIC) has been 
studying these topics experimentally since 2000, and was joined in 
November 2010 by the Large Hadron Collider (LHC) with  Pb+Pb collisions at an order of magnitude higher energy than the highest studied at RHIC. 

Since the first data were taken at RHIC, one of the most important experimental
signatures of the QGP has been the azimuthal anisotropy in correlations between detected
particles. In particular, the large value of the so-called `elliptic flow' observable, indicating strong collective behavior of the collision system, has been one of the most important and most studied measurements. It provided one of the strongest pieces of evidence leading to the conclusion that a strongly-coupled, low-viscosity, QGP medium is created in these collisions.

Elliptic flow refers to a second Fourier component of the azimuthal distribution of emitted
particles. When two identical nuclei collide at a finite impact parameter, the overlap region is an oblong shape in the transverse plane. In the standard picture, the system comes to an approximate local equilibrium and expands according to (viscous) hydrodynamics. The
elliptic asymmetry in the initial state is transformed during the collective expansion into an asymmetry in the final momentum distribution of the detected particles. The efficiency of
this transformation is sensitive to medium properties such as viscosity.

Random variations in the initial distribution of nucleons, due to quantum fluctuations, lead to event-by-event fluctuations in the initial geometry. Naively, in a symmetric collision system, odd harmonics in the azimuthal momentum distribution are expected to be negligible. It has only recently been realized that these fluctuations are not, in fact, negligible, and that by taking them into account one might potentially explain all long-range pair correlation data, which were previously not understood. Further, with this realization come new potential `flow'
observables --- not only the natural extension of elliptic flow measurements to odd harmonics such as triangular flow, but also a large number of other correlations.

These new measurements have the potential to provide tight constraints on theory and to extract precise quantitative properties of the QGP, as well as to shed light on the (as-yet poorly understood) non-equilibrium QCD dynamics of the initial stage of the collision.
The overall aim of the workshop was to construct a consistent picture of the
current understanding of the community and set a course for future 
investigation. 
Specific questions that were addressed during this workshop include:

\begin{itemize}
\item 
What is the best model for the initial state? What constraints can we already put on it from new data?
\item
How exactly is the initial spatial anisotropy converted to final-state momentum
anisotropy? Can we understand the hydrodynamic response in general, or do we need
to run event-by-event hydrodynamic simulations for every candidate initial condition?
\item
What is the current uncertainty in the viscosity of the QGP, and what is the best
strategy for reducing it?
\item
What else can the newly-measured observables teach us, and what other 
observables should be measured?
\item
What are the prospects for experimentally and/or theoretically distinguishing between
initial-state and collective effects?
\item
Can one disentangle various sources of fluctuations?
\end{itemize}

In the following we summarize the presentations and discussions at the 
workshop keeping the above formulated set of questions as a guideline. 
In section \ref{ini_cond} the understanding of the initial state is 
described in more detail. Section \ref{hydro}, contains the present 
analysis of the hydrodynamic response and the extraction of transport 
coefficients. In section \ref{comp} the challenges in comparing 
theoretical calculations with experimental measurements in a meaningful 
fashion are highlighted. Finally, section \ref{future} outlines a set 
of new observables that will enhance our understanding of hot and dense 
QCD matter. Note that we do not aim to provide a comprehensive review, 
but only offer highlights.  The slides from all talks can be obtained 
from~\cite{trento_talks}. 
%This workshop was very timely, coming at the 
%confluence of major theoretical and experimental developments that 
%resulted in a high degree of productiveness and the generation of important new ideas that will be summarized in this document and are expected to be worked out during the coming few years.

\section{Initial Conditions}
\label{ini_cond}

There is no longer any question that 
fluctuations in the initial conditions are crucial for a complete 
understanding of the bulk properties in heavy ion collisions~\cite{Roland}. 
The most basic way to model these fluctuations is to choose random positions for nucleons in each nucleus before the collision, and deterministically calculate the resulting post-collision energy density according to some prescription.  This is what is done, e.g., in a standard Monte Carlo Glauber or Monte Carlo KLN model, which have been commonly used in the past.

Although it is possible that this captures a significant part of the fluctuations in the initial transverse density, it is not sufficient for the type of quantitative comparison to data that will be necessary.
Within the last two years, there has been tremendous progress in the 
development of more realistic initial state descriptions. A wealth of 
different incarnations of the aforementioned Glauber and KLN model 
have been studied, e.g. including the fluctuations in the energy 
deposition per binary collision \cite{Dumitru} or taking into account a realistic wounding nucleon profile and the
nucleon-nucleon correlation correlations \cite{Broniowskidisc, Alvioli}. In addition, many groups use 
dynamical transport approaches to describe the initial non-equilibrium 
evolution of the heavy ion reaction \cite{Ollitrault,Werner,Petersen,Pang}.  

With all of this work ongoing, it is important to emphasize that there 
is not a binary choice between two well understood competing models for 
the initial condition (e.g.,  'Glauber' versus 'CGC'), as is sometimes 
inferred by those not directly involved in this research, but instead 
there exists a large space of possible physical pictures, parameters, 
and implementations.  For example, the Color-Glass-Condensate 
framework is not synonymous with only the particular MC-KLN model, 
but many implementations exist of CGC-based, Glauber-based, and
other ideas.
As such it is also important for researchers to specify the initial state model they employ with all necessary details including all the sources of fluctuations that are taken into account. Once the ingredients are better understood and there is a consensus reached on how to implement the fluctuations and geometry, it would be beneficial for standardized versions of the models to be made available to the public. 

The following sources of fluctuations were studied here: 
\begin{itemize}
\item
Fluctuations in the positions of the nucleons (or quarks) and their binary interactions
\item
Finite extent of the nucleons and correspondingly adjusted Wood-Saxon distributions
\item
Fluctuations in the initial momentum distribution (initial flow)
\item
Local fluctuations in energy deposition/particle production 
\end{itemize}

One of the key points that has been discussed in detail at the workshop is that the multiplicity distributions in proton-proton collisions provide an important constraint for the latter source of fluctuations. 
There is no unambiguous mapping of the fluctuations in p-p collisions to systems with higher nucleon density, but any realistic model should be able to reproduce the multiplicity distribution in the p-p limit. 

Often, the discussion of the initial state profiles has been 
restricted to the transverse plane, but the matter produced in heavy 
ion collisions has three spatial directions. Since there are more and 
more 3+1 dimensional (viscous) hydrodynamic codes in use, it is 
definitely time to pay more attention to the longitudinal direction. 
There have been a few exploratory studies on longitudinal fluctuations~\cite{Gavin,Poskanzer}, but a comprehensive understanding is still missing. The $\eta$ and $\Delta\eta$ dependence of particle correlations can lead to useful insights in that respect. 

The initial non-equilibrium evolution is still one of the main open 
questions in the field. Recently, there was a new promising attempt 
using SU(3) Yang-Mills field evolution by the BNL group \cite{Schenke}. 
It will be particularly interesting to see if rapid thermalization can be obtained within this framework
by including quantum fluctuations in a full 3+1D simulation.

Since the initial evolution produces finite initial velocity
  fields, a complete energy-momentum tensor including shear stress
  components should be used to initialize the hydrodynamic evolution. There are a few studies on the influence of initial flow in the system, but a consensus on its size and importance has not been reached yet. 

At the moment, the primary means of characterizing initial state profiles is calculation of the first few coefficients of the Fourier expansion in coordinate space ($\epsilon_n$). It needs to be explored whether there are other quantities that are suitable for representing the longitudinal direction, the initial velocity profile and other features in a more complete way. 

Overall, the main task at hand is to constrain the scale of the fluctuations exemplified by flux-tube radius, gaussian width, and the amplitude of fluctuations or correlation length \cite{Muller} in connection with a specific physics assumption. Now that it has become clear that quantum fluctuations are important to understand the full evolution of heavy ion reactions, there is the opportunity to pin down the highly excited nuclear initial states and its properties.  
\section{Hydrodynamic Response}
\label{hydro}
After this initial stage of a heavy-ion collision, the system continues to evolve, expanding collectively in response to these initial conditions.  This evolution is usually modeled with hydrodynamics or transport calculations, many of which were presented at this workshop.  Several important themes emerged from numerical simulations as well as  analytic work.

One emerging theme was a more comprehensive theoretical study of experimental data.  In the past, a few observables were commonly calculated and studied individually.  In the future it will be necessary to describe multiple observables from a single calculation, as well as to do so in more detail.
At a minimum, a simultaneous description of multiple flow harmonics 
$v_n\{2\}$ is needed as a function of transverse momentum and centrality, with 
even more information contained in higher cumulants~\cite{Murase}.  Especially challenging is a combined understanding of hard and soft physics, allowing for a description over a large range in transverse momentum~\cite{Werner}.  

However,  measurements that are even more differential are also possible (and available).
For example, a measurement like $v_2\{2\}$ is obtained by measuring azimuthal correlations between pairs of particles, but a single such measurement typically contains an average over the mean and relative pseudorapidity of the pair (within some range), as well as the transverse momentum of one or both particles, and disregarding information about other properties such as electric charge.  Nevertheless, more detailed dihadron correlation data are available to be studied.

In this workshop we saw progress in understanding how structures in relative pseudorapidity can arise from longitudinal fluctuations in the initial state~\cite{Pang, Gavin, Ko}, from intrinsic fluctuations generated during hydrodynamic evolution~\cite{Stephanov}, or from charge balancing that result in a differing structure for like-sign and opposite-sign pair correlations~\cite{Broniowski, Bozek}. 
In addition, progress was presented in the study of fluctuations in multiplicity and transverse momentum~\cite{Bozek, Moschelli}.
Most of these studies focussed on charged hadrons, but electromagnetic probes also give valuable independent information~\cite{Gale}.

Looking beyond two-particle correlations allows for a significantly expanded 
space of independent of observables.  One particularly exciting 
development was the large set of recent measurements of correlations between mixed harmonic event planes from the ATLAS collaboration~\cite{Mohapatra}, compared to brand new event-by-event hydrodynamic calculations~\cite{Heinz}.

Motivated by the large uncertainties that still remain in the initial conditions, another emerging theme is the characterization of the medium response in a general way.
A more detailed understanding of hydrodynamic response to initial 
conditions, as well as a precise determination of which aspects of the 
initial conditions each observable is most sensitive to, would allow 
for significant constraints to be placed on the properties of the 
initial conditions~\cite{Retinskaya} and medium 
properties~\cite{Luzum}. Exploring the scaling properties of flow 
observables may also help to understand system properties~\cite{Lacey}.  

Event-by-event hydrodynamic calculations have suggested that $v_2$ and $v_3$ can be accurately predicted in any given event by an eccentricity $\varepsilon_2$ and triangularity $\varepsilon_3$ in the initial transverse energy density profile, while $v_4$ and $v_5$ do not follow such a simple relation~\cite{Qiu}. This observation motivated a quantitative study of which definitions of $\varepsilon_2$ and $\varepsilon_3$ are the best predictors of final-state anisotropy, as well as a result that $v_4$ ($v_5$) arises as a  linear combination of $\varepsilon_4$ and $\varepsilon_2^{\ 2}$ ($\varepsilon_5$ and $\varepsilon_2\varepsilon_3$)~\cite{Ollitrault}.  This knowledge can be exploited by simulating only as many events as necessary to calculate the coefficient in front of each term, which allows systematic study of the importance of each term as a function of, e.g., $p_t$ and viscosity. Additionally, various predictions are made possible without the use of full event-by-event calculations~\cite{Teaney}. An open question remains as to how far this can be pushed --- e.g., can one reproduce all of the mixed harmonic event plane correlations without resorting to brute force event-by-event calculations?

\section{Comparison of Theory Calculations to Experimental Data}
\label{comp}
There has been a wealth of new measurements of higher anisotropy 
coefficients and other particle correlation observables presented at 
the workshop, e.g \cite{Mohapatra, Bilandzic, Li, Selyuzhenkov}. 
Some questions arose as to how a meaningful comparison to theory calculations can 
and should be performed, which is crucial for any quantitative 
statements about the transport coefficients of the quark gluon plasma.  

The first issue to be addressed is the definition and selection of centrality categories.  Theoretical calculations are often carried out at a specific impact parameter or in a range of impact parameters. This quantity is not directly accessible experimentally, but centrality classes are defined by the number of charged particles produced at mid- or forward rapidity, or by the number of spectators measured in a veto calorimeter. Usually the parameters in initial state models are tuned such as to reproduce the measured number of final-state particles in central collisions.  Parton-hadron duality is sometimes assumed in order to relate the initial number of gluons to the final number of pions. This mapping procedure needs to be carefully cross-checked, especially when the evolution is dissipative and the entropy increases. Differing centrality selection criteria can significantly influence the results, and any assumptions relating the impact parameter to the number of gluons and the final particle yields needs to be verified. A conclusion from the workshop was that the centrality determination procedure should be stated clearly when results are presented.

The second emphasis of the discussions at the workshop was more 
specifically related to flow observables. For an apples-to-apples 
comparison it is important to be aware of the different measurement 
methods and what they actually imply. In pure hydrodynamic 
calculations where observables are obtained by integrating over the 
Cooper-Frye hypersurface, one can calculate an exact value for $v_n$
in each event and therefore any possible event-averaged moment, e.g., the root mean 
square (RMS) value which can be compared to a 2-particle correlation measurement of $v_n$.
If a finite number of particles are sampled on the 
hypersurface that are potentially propagated through a hadron cascade, 
one must use procedures analogous to experimental analyses, such as 
the scalar product method, 2-particle correlation 
method or the cumulant method which can be compared directly to the relevant data. 
It is important to note, however, that the standard event plane method only gives 
the same result as the corresponding experimental measurement
if the resolution of the event plane is the same in both cases.

Keeping in mind caveats such as this, the best way to make a meaningful comparison between 
theoretical calculations and experimental data is to employ the same 
analysis as used by the experiments. The first step is to define the centrality classes in the same way as the specific 
collaboration. The MIT group (CMS) and Ante Bilandzic (ALICE) have 
expressed interest in collaborating with theoreticians to provide 
them with stand-alone software which can be used for the analysis of final 
state particle distributions. The first package with the cumulant 
analysis developed by the ALICE collaboration is currently being 
tested. 
\section{Outline of Future Measurements}
\label{future}
One of the discussion sessions during the workshop has been devoted to 
future experimental measurements and requests to the theory community. 
The extensive factorization tests of two-particle correlations ($V_{n\Delta}$) over a large range of 
trigger and associated particle transverse momentum as carried out by 
the LHC collaborations are useful 
to distinguish bulk anisotropies from other/non-flow contributions,
as well as to study flow fluctuations.  Similar analyses can be done
in the space of trigger and associated pseudorapidity.
Measurements of $v_n$ coefficients with respect to an event plane of 
different order $\Psi_m$ with $m\neq n$ are sensitive to the hydrodynamic 
response and give insights about mode-mixing. Reconstructing the 
$\Psi_1$ event plane that corresponds to the rapidity-even $v_1$ 
observable would be desirable. Another measurement that could provide 
insights into the interplay of jet-medium response and pure bulk 
evolution effects is to analyse the untriggered two-particle 
correlations in events that contain a 50-100 GeV jet and compare the 
result to the minimum bias one. One issue that might complicate such a 
measurement is to find a consistent centrality selection, since the 
requirement of a high momentum jet biases the event sample. 

The wish-list for the theory community is to calculate for example 
3-particle correlations with a high transverse momentum ($6-8$ GeV) 
trigger particle and associated particles below $\sim 2$ GeV in all 
different combinations of angular spaces, i.e. $\Delta \phi-\Delta 
\phi$, $\Delta \phi-\Delta \eta$ and $\Delta\eta-\Delta\eta$. As a 
baseline the correlation functions from a medium evolution only would 
be interesting already. Otherwise, these multi-particle calculations 
are the most promising way to disentangle jet-medium effects from the 
underlying pure medium response, if such a distinction makes sense. 
The meaning of the event-plane correlations as measured by the ATLAS 
collaboration should be investigated further by more theory 
comparisons. In addition, more theoretical effort needs to be spent on 
calculating other fluctuation observables, such as $\langle p_T 
\rangle$ fluctuations, that have sensitivity to the number of sources in 
the initial state. Not all theory groups are able to address these types of 
observables, since very high statistics are required which corresponds 
to a huge amount of CPU time in event-by-event hybrid approaches. 
Another goal that might be easier to reach is to perform calculations 
of higher harmonic flow coefficients at lower beam energies, since there are 
now results available from the low beam energy scan at RHIC, as well as for other nuclei (copper and uranium). The flow 
results for identified particles should also be addressed by more 
theoretical calculations. 

\section[conclusions]{Conclusions and Outlook}
\label{concl}
All known sources of fluctuations should be included in models of the initial stages of
heavy-ion collisions. In particular, implementations of fluctuations in particle production into
Monte Carlo models should be continued, making sure to obey experimental constraints like
multiplicity distributions in proton-proton collisions, and differences in recent implementations
should be better understood.
Much progress has been made in characterizing the hydrodynamic response to the
initial conditions in terms of simple relationships between properties of the initial density and
flow correlations in the final particle distributions. This is providing us with important insight, but
it is an open question how far this can be taken and whether brute-force event-by-event
calculations will always be necessary for describing certain data.
More standardization is necessary in the field. Examples include definitions of initial
anisotropies $\epsilon_n$, variations of Glauber Monte Carlo models, and experimental flow
analyses.
Subtleties are present in comparing theoretical calculations to experimental data and
more care needs to be taken in the future in order to compare the correct quantities.
Some possibilities for future measurements and a wish-list for the 
theory community have been presented in Section~\ref{future}. 
%The field of initial state fluctuations and final state correlations 
%has seen a rapid development in the last 2 1/2 years, and we expect that 
%the questions outlined in this note will be addressed in the 
%near future. 

This workshop was very timely, coming at the 
confluence of major theoretical and experimental developments  
resulting in a high degree of productiveness and in the generation 
of important new ideas summarized here,
which we expect to be worked out during the coming few years.

\section*{Acknowledgements}
We are grateful to all the participants of the Trento workshop for 
excellent presentations of their work and enlightening discussions.   
H.P is supported by U.S. department of Energy grant DE-FG02-05ER41367 
and acknowledes funding of a Helmholtz Young Investigator Group 
VH-NG-822. 
M.L.~was funded by
the European 
Research Council under the Advanced Investigator Grant ERC-AD-267258
and by the Natural Sciences and Engineering Research Council 
of Canada.

\end{document}